\documentclass[twocolumn, a4paper, 11pt,book]{archive}
\usepackage{version}
\usepackage{graphicx}

\usepackage[]{natbib}
\usepackage{txfonts}

\usepackage{booktabs}
\usepackage{multirow}

\usepackage{rotating} % Rotating table
\usepackage{color, colortbl}

\usepackage{siunitx}

\definecolor{Gray}{gray}{0.9}
\definecolor{LightRed}{rgb}{1., 0.88, 0.88}
\definecolor{LightCyan}{rgb}{0.88,1,1}
\definecolor{LightGreen}{rgb}{0.8,0.98,0.8}
\definecolor{LightViolet}{rgb}{0.95, 0.85, 1.0}
\definecolor{azur}{rgb}{0.94, 1.0, 1.0}

\usepackage{draftwatermark}
\SetWatermarkText{DRAFT}
\SetWatermarkScale{0.75}
\begin{document} 

\twocolumn[{%
 \centering
%
  %\title{}
{\center \bf \Huge Mechanochemical synthesis of Aromatic Infrared Band carriers.}\\
\vspace*{0.05cm}
%{\center \bf \Huge }\\
%\vspace*{0.05cm}
{\center \bf \Large The top-down chemistry of interstellar carbonaceous dust grain analogues}\\
\vspace*{0.25cm}

{\Large E. Dartois \inst{1},
        E. Charon\inst{2},
        C. Engrand\inst{3},
        T. Pino\inst{1},
        C. Sandt\inst{4}
       }\\
\vspace*{0.25cm}

$^1$      Institut des Sciences Mol\'eculaires d'Orsay, CNRS, Universit\'e Paris-Saclay, 
B\^at 520, Rue Andr\'e Rivi\`ere, 91405 Orsay, France\\
              \email{emmanuel.dartois@universite-paris-saclay.fr}\\
$^2$                  NIMBE, CEA, CNRS, Universit\'e Paris-Saclay, CEA Saclay 91191 Gif-sur-Yvette France\\
$^3$                 IJCLab, CNRS/IN2P3, Universit\'e Paris-Saclay, 91405 Orsay, France\\
 $^4$                Synchrotron Soleil, L'Orme des Merisiers, BP 48 Saint Aubin, 91192, Gif-sur-Yvette Cedex, France\\
 \vspace*{0.5cm}
{keywords: AIB, PAH, Astrochemistry, ISM: lines and bands, ISM: dust, extinction, ISM: planetary nebulae: general, Infrared: ISM}\\
 \vspace*{0.5cm}
{\it \large To appear in Astronomy \& Astrophysics}\\
%
%    \thanks{Part of the equipment used in this work has been financed by the French INSU-CNRS program``Physique et Chimie du Milieu Interstellaire'' (PCMI).}
%
% \abstract{}{}{}{}{} 
% 5 {} token are mandatory
%  \vspace*{0.5cm} 
 }]
  \section*{Abstract}
   {
    Interstellar space hosts nanometre- to micron-sized dust grains, which are responsible for the reddening of stars in the visible.
The carbonaceous-rich component of these grain populations emits in infrared bands that have been observed remotely for decades with telescopes and satellites. 
 They are a key ingredient of Galactic radiative transfer models and astrochemical dust evolution.
 However, except for C$_{60}$ and its cation, the precise carriers for most of these bands are still unknown and not well reproduced in the laboratory.
}
  % aims heading (mandatory)
   {  
   In this work, we aim to show the high-energy mechanochemical synthesis of disordered aromatic and aliphatic analogues provides interstellar relevant dust particles.%\LEt{I'm sorry but I really can't work out what you're trying to say here. Please review for clarity or send me an explanation if necessary. Also, please do be cautious when using the slash. This sign is used first and foremost to indicate a ratio and secondly to replace "or". Please check for this throughout to avoid misinterpretation. For details, refer to Sect. 3.6 of Language Guide. (https://www.aanda.org/for-authors/language-editing/1-introduction)  }
}
  % methods heading (mandatory)
   { 
   The mechanochemical milling of carbon-based solids under a hydrogen atmosphere produces particles with a pertinent spectroscopic match to astrophysical observations of aromatic infrared band (AIB) emission, linked to the so-called astrophysical polycyclic aromatic hydrocarbon (PAH) hypothesis. 
The H/C ratio for the analogues that best reproduce these astronomical infrared observations lies in the 5$\pm{2}$\% range, potentially setting a constraint on astrophysical models. This value happens to be much lower than diffuse interstellar hydrogenated amorphous carbons, another Galactic dust grain component observed in absorption, and it most probably provides a constraint on the hydrogenation degree of the most aromatic carbonaceous dust grain carriers. A broad band, observed in AIBs, evolving in the 1350-1200 cm$^{-1}$ (7.4-8.3~$\mu$m) range is correlated to the hydrogen content, and thus the structural evolution in the analogues produced.
}
  % results heading (mandatory)
   { 
 Our results demonstrate that the mechanochemical process, which does not take place in space, can be seen as an experimental reactor to stimulate very local energetic chemical reactions. It introduces bond disorder and hydrogen chemical attachment on the produced defects, with a net effect similar to the 
 interstellar space very localised chemical reactions with solids.
 %\LEt{very long sentence. Unclear what it means at the end. Perhaps you could split it into 2 sentences in order to make it clearer?}
From the vantage point of astrophysics, these laboratory interstellar dust analogues will be used to predict dust grain evolution under simulated interstellar conditions, including harsh radiative environments. 
Such interstellar analogues offer an opportunity to derive a global view on the cycling of matter in other star forming systems.
}
    % conclusions heading (optional), leave it empty if necessary 
%   {}
%
%
%   \maketitle
%
%________________________________________________________________

\section{Introduction}
%
%-------------------------------------------------------------------------------------------------------------------------------------------------------------------
\begin{table*}%[htdp]
%\begin{sidewaystable*}
\caption{Observation summary}
\begin{center}
\begin{tabular}{l l l  l l l   }
\hline
\rowcolor{azur} Object  &Right Ascension &Declination &Obs. identifier or reference &Class &Type (Simbad)\\
\hline
SMC 2MJ004441   &00h 44m 41.05s &-73d 21' 36.4"         &Spitzer AORKEY 27525120                &$\mathcal{D}$          &Protoplanetary Nebula          \\
IRAS 04296+3429 &04h 32m 56.976s        &+34d 36' 12.51"        &Spitzer AORKEY 4116736          &$\mathcal{C,D}$                        &Protoplanetary Nebula  \\
                                &                               &                               &UKIRT; \cite{Geballe1992}              &       &                       \\
IRAS 05073-6752 &05h 07m 13.920s        &-67d 48' 46.73"        &Spitzer AORKEY 24317184         &$\mathcal{D}$                  &Protoplanetary Nebula  \\
IRAS 05110-6616 &05h 11m 10.639s        &-66d 12' 53.63"        &Spitzer AORKEY 25992704         &$\mathcal{D}$                  &Post AGB       \\
IRAS S05211-6926        &05h 20m 43.58s &-69d 23' 41.4"         &Spitzer AORKEY 27985920         &$\mathcal{D}$                  &Pulsating Star \\
IRAS Z05259-7052        &05h 25m 20.75s &-70d 50' 07.3"         &Spitzer AORKEY 25996032         &$\mathcal{D}$                  &Variable Star  \\
IRAS 05289-6617         &05h 29m 02.412s        &-66d 15' 27.77"        &Spitzer AORKEY 11239680         &$\mathcal{D}$                  &Carbon star    \\
IRAS 05341+0852 &05h 36m 55.050s        &+08d 54' 08.65"        &Spitzer AORKEY 4896512          &$\mathcal{D}$                  &Protoplanetary Nebula  \\
                                &                               &                               &UKIRT; \cite{Joblin1996}                       &       &                       \\
IRAS 06111-7023 &06h 10m 32.008s        &-70d 24' 40.72"        &Spitzer AORKEY 19013120         &$\mathcal{D}$                  &Star           \\
IRAS 13416-6243 &13h 45m 07.65s         &-62d 58' 19.0"         &ISO TDT 62803904                                &$\mathcal{C}$          &Protoplanetary Nebula          \\
IRAS 15038-5533 &15h 07m 34.79s         &-55d 44' 50.8"         &Spitzer AORKEY 17936384                 &$\mathcal{D}$                  &Far-IR source                 \\
IRAS 20000+3239 &20h 01m 59.513s        &+32d 47' 32.81"        &ISO TDT 18500531                                &$\mathcal{D}$                  &Variable Star    \\
                                &                               &                               &Spitzer AORKEY 4897792          &       &                       \\
AFGL 2688               &21h 02m 18.79s         &+36d 41' 37.4" &ISO TDT 35102563                                &$\mathcal{C}$          &Protoplanetary Nebula  \\
                                &                               &                               &UKIRT; \cite{Geballe1992}              &       &                       \\
IRAS 22272+5435 &22h 29m 10.375s        &+54d 51' 06.34"        &ISO TDT 26302115                                &$\mathcal{D}$                  &Protoplanetary Nebula  \\
                                &                               &                                &Subaru; \cite{Goto2003}                        &       &                       \\
IRAS 23304+6147         &23h 32m 44.785s        &+62d 03' 49.08"        &ISO TDT 39601867                            &$\mathcal{D}$                  &Protoplanetary Nebula  \\
                                &                               &                               &Spitzer AORKEY 14401024         &       &                       \\
\hline
\end{tabular}
\end{center}
\label{Table_obs_summary}
\end{table*}%
%-------------------------------------------------------------------------------------------------------------------------------------------------------------------
%
Our Galaxy hosts carbonaceous dust grains displaying emission and absorption bands observed by telescopes in the infrared.
The emission in the diffuse interstellar medium of these bands, dominated by an aromatic vibrational character, also called AIBs (aromatic infrared bands), has led to the so-called PAH
%\LEt{Please ensure & check throughout the paper that all acronyms and abbreviations are written out at first mention, followed by the abbreviation or acronym in parentheses (even if you have already introduced them in the Abstract). After that please use only the abbreviation. Instruments, surveys, or facilities do not need an introduction in the Abstract. Please introduce these in the body of the paper unless they are known only by their acronym. See Sect. 5.2.4 of the Language Guide (https://www.aanda.org/for-authors/language-editing/1-introduction) } 
hypothesis. Under this theory, the observed emission is related to the infrared fluorescence emission mechanism of polycyclic aromatic hydrocarbon-like molecules \citep{Leger1984, Allamandola1985}, following energetic photon absorption, although no unique PAH has been identified in the mid-infrared so far.
Apart from the recent attribution of a few specific infrared emission bands to the C$_{60}$ and possibly C$_{70}$ fullerene molecules in some sources \citep{Sellgren2009, Sellgren2010, Cami2010}, the carriers of the AIBs remain elusive.
The emission bands have been categorised in different classes ($\mathcal{A}$ to $\mathcal{D}$) following the ascertainment of band profiles and center position variations, ensuing from a phenomenological deconvolution of the observations.
 \citep{Peeters2002, vanDiedenhoven2004, Matsuura2007, Sloan2007, Keller2008, Boersma2008, Pino2008, Acke2010, Carpentier2012, Gadallah2013}.
In the late classes of infrared emission spectra (so-called $\mathcal{C}$ and $\mathcal{D}$), a mix between an aromatic and aliphatic character is observed.
Class $\mathcal{A}$ sources are dominated by aromatic bands, whereas classes $\mathcal{C}$ and $\mathcal{D}$ harbour a pronounced aliphatic character. 
Most of the interstellar hydrocarbons are injected through the late phases of stellar evolution, post asymptotic giant branch (AGB) and protoplanetary nebula (PPN), which often display the class $\mathcal{C}$ and $\mathcal{D}$ spectral character.
In absorption, bands at 3.4, 6.85, and 7.25 microns are also observed in the diffuse interstellar medium (ISM) of our Galaxy, as well as in extragalactic ISM, and can be well represented by a material with a significant amount of aliphatic character, also called HAC or a-C:H, which is a family of hydrogenated amorphous carbons 
\citep[e.g.][ and references therein]{Allen1981, Duley1983, Mason2004, Risaliti2006, Dartois2007, Dartois2007b, Imanishi2010}.
However, the AIB emission spectra are not a simple linear combination of aliphatic (such as the a-C:H observed in absorption in the diffuse ISM) and aromatic (class $\mathcal{A}$ observed in emission) spectra. They show a spectral evolution of vibrational modes, including the shift of the C=C mode from 6.2 to 6.3 microns from class $\mathcal{A}$ to $\mathcal{C}$ \citep[e.g.][]{vanDiedenhoven2004}, and broadening and shifting back to lower wavelengths in class $\mathcal{D}$. Unlike classes $\mathcal{A-B}$ with two bands in the 7.6 to 8.2 microns range, classes $\mathcal{C-D}$ show a broad band
 \citep[][]{Szczerba2005,Matsuura2014}, with the class $\mathcal{D}$ one peaking at a shorter wavelength. The out of plane, predominantly aromatic, CH vibration patterns (from 11 to 13 $\mu$m) are generally difficult to reproduce with dust analogues in the laboratory. 
In addition to chemically pure PAH studies, a large range of dust grain analogues have been tailored using lasers, flames, VUV continuum sources, and/or plasmas in the laboratory to tackle this identification issue (Carpentier et al. 2012, Schnaiter et al. 1999, Jager et al. 2006, Mennella et al. 1999, Dartois et al. 2005, Furton et al. 1999, Biennier et al. 2009).
Among the issues faced by laboratory synthesis of interstellar dust analogues is the ability to produce an environment that is not homogeneous for the entire batch of analogues produced (i.e. localised modifications), as many processes in the interstellar medium will only affect grain (surfaces) moieties considering, for example, hydrogen or radical accretion/addition, cosmic ray impact,  grain-grain shocks, etc. As a consequence, a large inhomogeneity at the nanometre scale between adjacent constitutive elements making the grain may be expected, introducing many local defects that will influence the nature and spectroscopic properties of such small dust grains.
In this study, we address a new non-homogeneous, non-bottom-up shock approach. We used a mechanochemical synthesis method under a pressurised hydrogen atmosphere, to produce laboratory interstellar dust grain analogues to explain the infrared emission spectra of the remotely observed aliphatic and aromatic mixed interstellar dust grains observed through infrared emission bands.
%
%-------------------------------------------------------------------------------------------------------------------------------------------------------------------
%
\begin{table*}%[htdp]
%\begin{sidewaystable*}
\caption{Experiment summary}
\begin{center}
\begin{tabular}{l l l  l l l  l l l  }
\hline
Exp.    &Precursor (weight) &BPR$^a$ &H$_2$$^b$ &RPM$^c$ &Time$^d$       &H/C    &HRTEM$^e$ &\\
        & & &(bar) &     & &(at. \%)            & &\\
\hline
\rowcolor{LightRed} \multicolumn{9}{l}{Stainless Steel}\\
\hline
sst-a   &Graphite (1.3 g)       &94     &3.5            &400            &90h            &4.27                 &- &                     \\
sst-b   &Graphite (1.7 g)       &72     &4.3            &400            &43h30         &3.52    &-                     & \\
sst-c   &Mesoporous C (5g)      &25     &13             &400            &44h30         &-                              &-                     & \\
sst-d   &Mesoporous C (5g)      &25     &16             &400            &45h45         &-                              &-                     & \\
sst-e$^{f}$     &Mesoporous C (5g)      &25     &16             &400            &45h45  &-       &-                     & \\
sst-f           &Graphite (2.5g)                &49     &20     &400            &71h30   &6.38    &y                     & \\
sst-g           &Fullerene (2.5g)               &49     &20     &400            &100h   &5.34    &y                     & \\
sst-h           &MWCNT (2.5g)           &46     &25     &400            &155h   & -                       &y                     & \\
sst-i1  &Graphite (1.33g)               &50     &20 &1500       &0h30   &1.12    &-                     & \\
sst-i2  &Graphite (1.33g)               &50     &20 &1500       &2h             &2.85    &-                     & \\
sst-i3  &Graphite (1.33g)               &50     &20 &1500       &8h             &3.63    &-                     & \\
sst-j1  &Graphite (1.57g)               &53     &20 &1500       &0h30   &2.48    &-                     & \\
sst-j2  &Graphite (1.57g)               &53     &20 &1500       &2h             &4.59    &-                     & \\
sst-j3  &Graphite (1.57g)               &53     &20 &1500       &8h             &4.41    &-                     & \\
sst-k           &Fullerene (1.33g)              &50     &20 &1500       &9h             &3.00    &-                     & \\
sst-l1  &MWCNT (1.31g)          &51     &20 &1500       &1h             &1.82    &-                     & \\
sst-l2  &MWCNT (1.31g)          &51     &20 &1500       &3h             &3.19    &-                     & \\
sst-l3  &MWCNT (1.31g)          &51     &20 &1500       &9h             &3.41    &-                     & \\
sst-m   &Carbon Black (1.32g)   &51     &20 &1500       &9h             &-                       &-                     & \\
\hline
\rowcolor{LightCyan} \multicolumn{9}{l}{Hardened Steel}\\
\hline
hst-a1  &Graphite (7.5g)                &83     &23     &400    &5h             &2.83    &y                     & \\
hst-a2  &Graphite (7.5g)                &83     &23     &400    &17h    &5.13    &y                     & \\
hst-a3  &Graphite (7.5g)                &83     &23     &400    &48h    &4.42    &y                     & \\
hst-b   &Graphite (1g)          &132            &20     &400 &8h &-      &-                     & \\
hst-c   &Graphite (18g)         &37     &20     &400 &72h &-     &-                    & \\
hst-d   &NanoGraphite (15g)     &14     &25     &400 &234h &-    &-                     & \\
hst-e   &Fullerene (5g)         &64     &25     &400 &100h &4.00         &y                     & \\
hst-f   &Carbon Black (5g)      &81     &17     &400 &96h &-     &y                     & \\
\hline
\rowcolor{LightGreen} \multicolumn{9}{l}{Tungsten Carbide}\\
\hline
cw-a1   &Graphite (5g)  &31     &20     &400            &8h &1.69        &y                     & \\
cw-a2   &Graphite (5g)  &31     &20     &400            &20h &2.38       &y                     & \\
cw-a3   &Graphite (5g)  &31     &20     &400            &100h &3.93      &y                     & \\
cw-b    &MWCNT (7.5g)   &70     &25     &400            &78h &- &-       &               \\
cw-c    &Fullerene (3g)         &174            &20     &400            &26h &-      &y                     & \\
\hline
\rowcolor{LightViolet}\multicolumn{9}{l}{Zirconium Oxide}\\
\hline
zr-a1           &Graphite (2.5g)        &31     &23     &400            &4h &-       &-                     & \\
zr-a2           &Graphite (2.5g)        &31     &23     &400            &24h &-       &-                     & \\
zr-a3           &Graphite (2.5g)        &31     &23     &400            &131h &3.55   &y                     & \\
zr-b            &Fullerene (1.1g)       &70     &23     &500            &91h &3.00    &y                     & \\
\hline
\end{tabular}
\end{center}
$^a$ Ball-to-powder ratio, i.e. weight of the milling balls with respect to the milled sample; $^b$ input pressure before milling; $^c$ revolutions per minute; $^d$ milling time. $^e$ High-Resolution Transmission Electron Microscope $^f$ HCl washed after milling.\\
\label{Table_summary}
\end{table*}%
%-------------------------------------------------------------------------------------------------------------------------------------------------------------------

%~~~~~~~~~~~~~~~~~~~~~~~~~~~~~~~~~~~~~~~~~~~~~~~~~~~~~~~~~~~~~~~~~~~~~~~~~~~~~~~~~~~~~~~~~~~~~~~~~~~~~~~~~~~~     
%
%       FIGURE 1
%
\begin{figure}%[tbhp]
\centering
\includegraphics[width=1.0\linewidth]{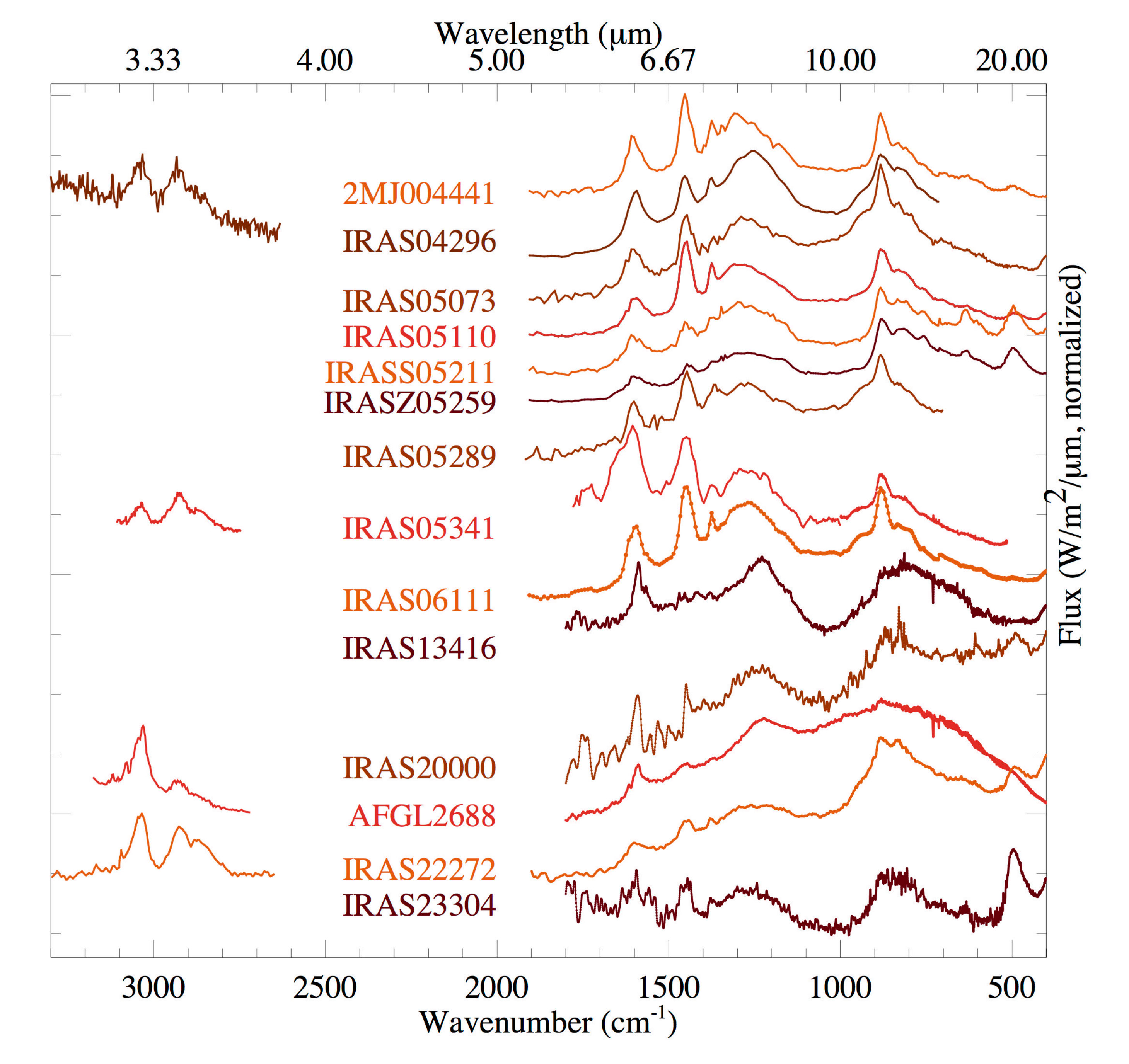}
\caption{Astronomical observations of class $\mathcal{C}$ and $\mathcal{D}$ sources, displaying characteristic infrared emission bands.
%\LEt{not sure what exactly is being emitted here? characteristic infrared-emitting bands? maybe something else?  }
}
\label{fig:obs}
\end{figure}
\section{Observations}%\LEt{Please could you number your sections throughout}}
Astronomical spectra were retrieved from the ISO\footnote{https://www.cosmos.esa.int/web/iso/access-the-archive} and Spitzer\footnote{http://archive.spitzer.caltech.edu/ and https://cassis.sirtf.com/} satellite databases, after a selection from the published literature on clear detections of $\mathcal{C}$ and $\mathcal{D}$ classes \citep{Matsuura2014, Sloan2007, Szczerba2005}. The existing ground-based L-band spectra (in the 2600-3300 cm$^{-1}$ range) were taken from the literature for IRAS04296-3429 and AFGL2688 \citep{Geballe1992}, IRAS 05341+0852 \citep{Joblin1996}, and IRAS 22272+5435 \citep{Goto2003} to complement the spectra at lower wavelengths. A summary of the targets is given in Table~\ref{Table_obs_summary}. The observed spectra are shown in Fig.\ref{fig:obs}, each one normalised by a scaling factor for clarity.
%
%~~~~~~~~~~~~~~~~~~~~~~~~~~~~~~~~~~~~~~~~~~~~~~~~~~~~~~~~~~~~~~~~~~~~~~~~~~~~~~~~~~~~~~~~~~~~~~~~~~~~~~~~~~~~     
%
%       FIGURE 0
%
\begin{figure*}%[tbhp]
\centering
\includegraphics[width=1.0\linewidth]{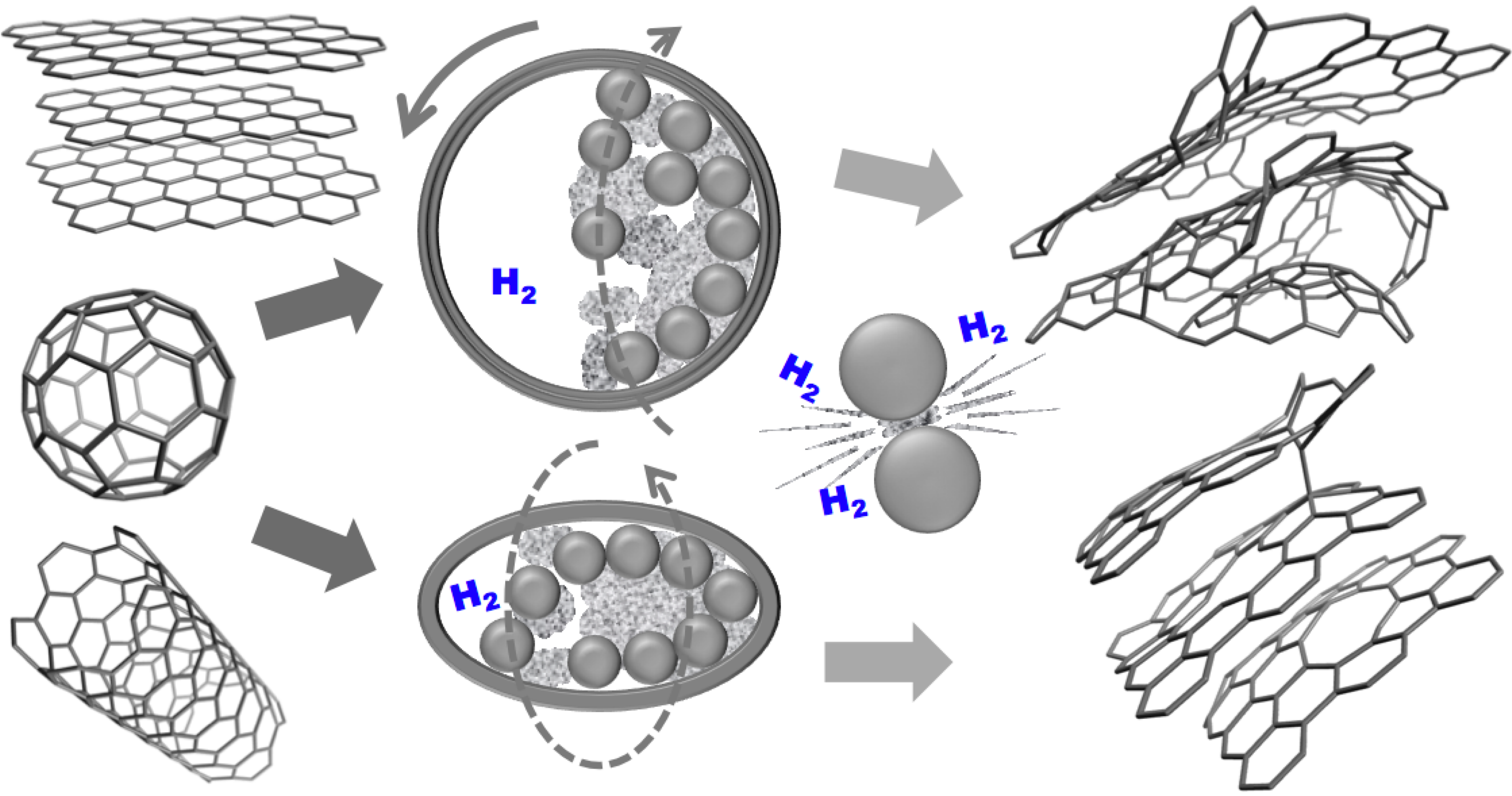}
\caption{Schematic representation of the top down mechanochemical process performed with a planetary ball mill (upper) and high-energy mill (lower). The milling impacts and shattering modify the size or crosslink graphite grains, flakes, carbon nanotubes and fullerenes down to the nanometre scale, and a defect-induced chemistry takes place in the pressurised
%\LEt{what is reducing what? Could it be 'reduced-pressure'?}
 hydrogen reducing atmosphere.}
\label{fig:experience}
\end{figure*}
%
%~~~~~~~~~~~~~~~~~~~~~~~~~~~~~~~~~~~~~~~~~~~~~~~~~~~~~~~~~~~~~~~~~~~~~~~~~~~~~~~~~~~~~~~~~~~~~~~~~~~~~~~~~~~~     
\section{Experiments}%\LEt{Please number your sections \& sub-sections. Also add intros before breaking into sub-sections (even just one sentence is enough)}}
%~~~~~~~~~~~~~~~~~~~~~~~~~~~~~~~~~~~~~~~~~~~~~~~~~~~~~~~~~~~~~~~~~~~~~~~~~~~~~~~~~~~~~~~~~~~~~~~~~~~~~~~~~~~~     
%
Several experiments were conducted to produce analogues by mechanochemical milling and analyse them by spectroscopic means, electron microscopy and elemental analysis.
\subsection{Mechanochemical milling}
Different (hydrogen-free) carbonaceous solid precursors (99.995\% pure graphite or mesoporous activated carbon; Sigma Aldrich; NanoGraphite, Graphene Supermarket; fullerene C$_{60}$ 99.9+\%, purified, SES research; Multi-wall carbon nanotubes (MWCNT) 98\% pure < 8~nm outer diameter, carbon nanotubes plus; carbon black nanopowder, average diameter 13nm, PlasmaChem) were inserted in vacuum-tight bowls. The choice of different carbon precursors was made to test to what extent the initial structure influences the results.
The bowls were adapted by in-house connections to firstly be evacuated to a secondary vacuum ($\rm <10^{-4}$ mbar) and then pressurised with hydrogen gas before being placed in a Retsch PM100  planetary (250ml or 500ml bowls) or a Retsch high-energy Emax (50ml bowls) ball mill.
In the planetary ball mill, the bowl containing grinding balls made of the same material as the bowl is locked at an eccentric position on a rotating sun wheel.
The jar and the wheel rotate in opposite directions. The accelerated balls' impacts and frictions at high relative speeds transfer high energy to the solid particles within the bowl.
In the Emax ball mill, the grinding jar's oval geometry and higher rotating speeds transfer even higher energy to the particles. 
A sketch showing the milling process, bowl shapes, and movements is given in Fig.\ref{fig:experience}.
During milling, the bowl temperature never exceeded 80$\rm^o$C for the PM100 experiments and was controlled and kept below 50$\rm^o$C for the Emax one, ensuring the modifications could not be induced by a high thermostatic temperature.
After milling, the samples were extracted in an inflatable polyethylene glove box, evacuated and refilled with Ar or N$_2$, in order to minimise oxygen contamination. The samples were transferred and curated in sealed glass vials placed in an evacuated desiccator.
One of the main contaminations of a long-term milling process comes from the abrasion of the bowl itself, producing (nano-)particles that mix within the agglomerated carbonaceous dust particles.
The strategy, in order to properly evaluate the impact of the bowl media on the resultant dust particles produced, was to conduct experiments using, when available, different bowl materials and stiffnesses (stainless steel, hardened steel, tungsten carbide and zirconium oxide). The infrared spectra and electron images on the final products were compared. 
The balls-to-powder-ratio (BPR), meaning the ratio of the milling balls to sample mass, determining the number and energy of the impacts, varied between 14 and 174. The higher this ratio, the more processed the sample should be. However, it also significantly increases ball-to-ball impacts and shattering, and thus the level of milling medium contamination.
%\LEt{Does this mean the level of milling medium contamination is significantly increased? If I've misunderstood then please review as it isn't clear} 
A too low BPR does not process the sample efficiently. BPRs centred around 50 seemed a good compromise experimentally.
A summary of the performed experiments is given in Table~\ref{Table_summary}.
%
%       FIGURE 2
%
\begin{figure}%[tbhp]
\centering
\includegraphics[width=1.0\linewidth]{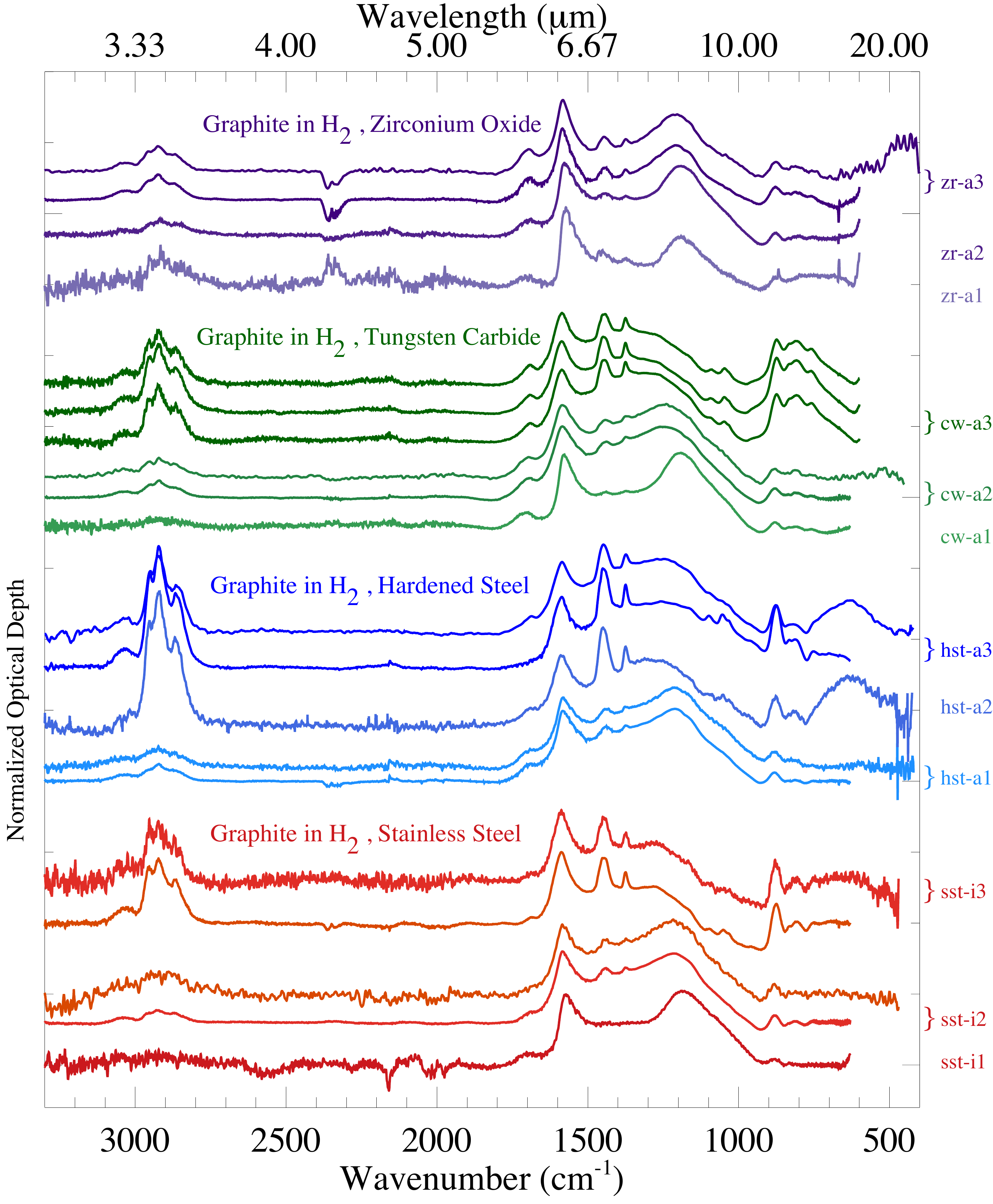}
\caption{Infrared spectra of a sample set of planetary ball-milled graphite
%\LEt{May be a need for hyphens but I risk changing the meaning. ex: planetary-ball milled graphite = milled graphite of the planetary ball, or planetary ball-milled graphite = graphite that has been ball milled, etc... please check here and throughout the paper \& ensure consistency where you do/do not agree with my suggestions     } 
under a pressurised hydrogen atmosphere in function of milling time. The millings were performed in bowls made of stainless steel, hardened steel, tungsten carbide, and zirconium oxide. The spectra were continuum subtracted.}
\label{fig:IR}
\end{figure}
%
%
%       FIGURE 4
%
\begin{figure*}%[tbhp]
\centering
\includegraphics[width=1.0\linewidth]{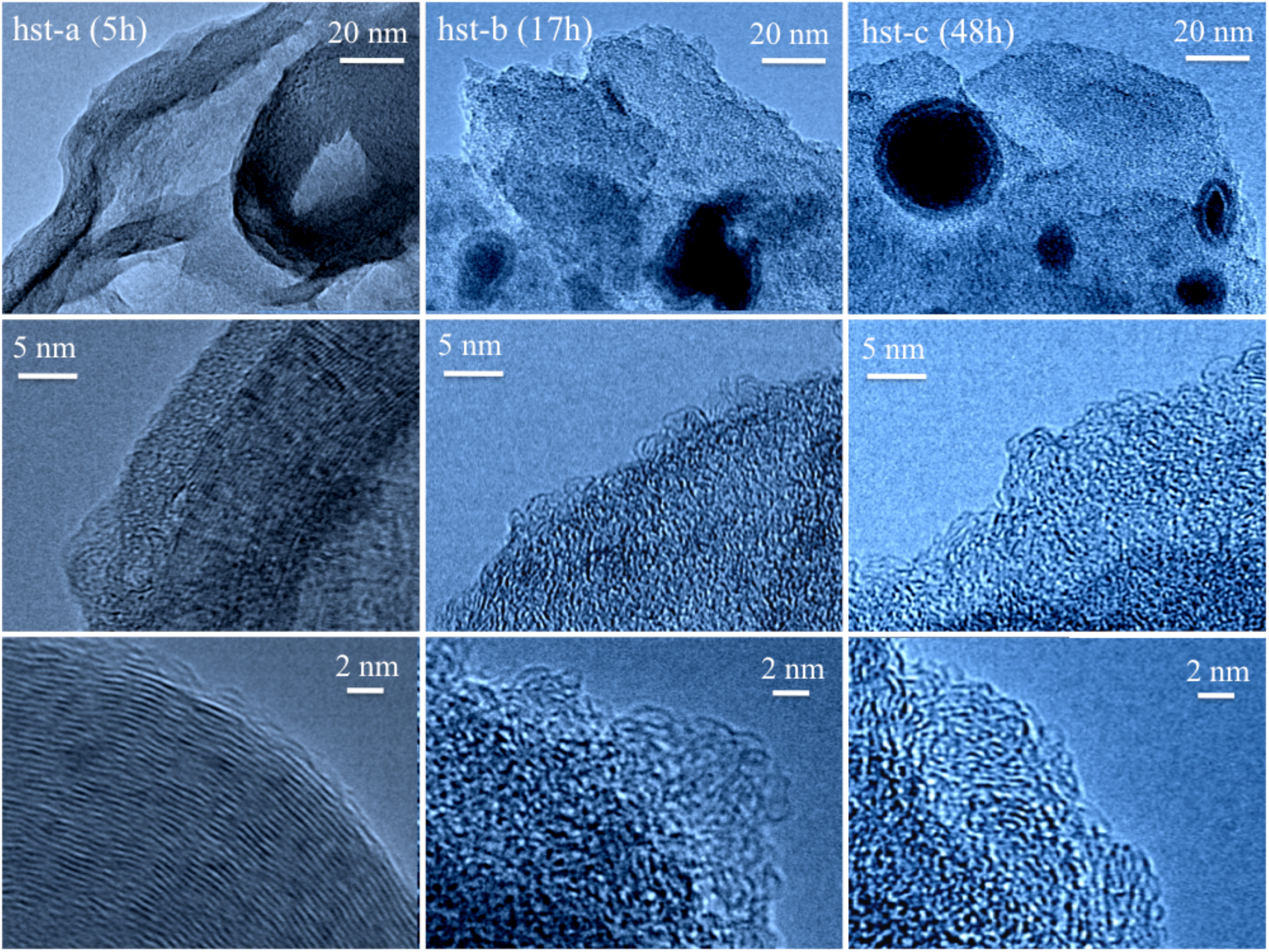}
\caption{High-resolution transmission electron microscope images at increasing magnification scales showing the structural evolution of ball-milled graphite under hydrogen atmosphere for the hardened steel experiment in a planetary ball mill after 5h (left panels), 17h (middle panel) and 48h (right panels). The ordered graphitic plane evolves rapidly toward a turbostratic structure, and, at high milling times, a fully disordered carbon network with a high curvature. We note the progressive contamination of the hydrogenated carbon structure by iron nanoparticles issued from the shattering of the bowl during the milling process, giving rise to dispersed isolated passivated iron nanoparticles.}
\label{fig:HRTEM}
\end{figure*}
%
%
%
%       FIGURE RAMAN
%
\begin{figure}%[tbhp]
\centering
\includegraphics[width=1.0\linewidth]{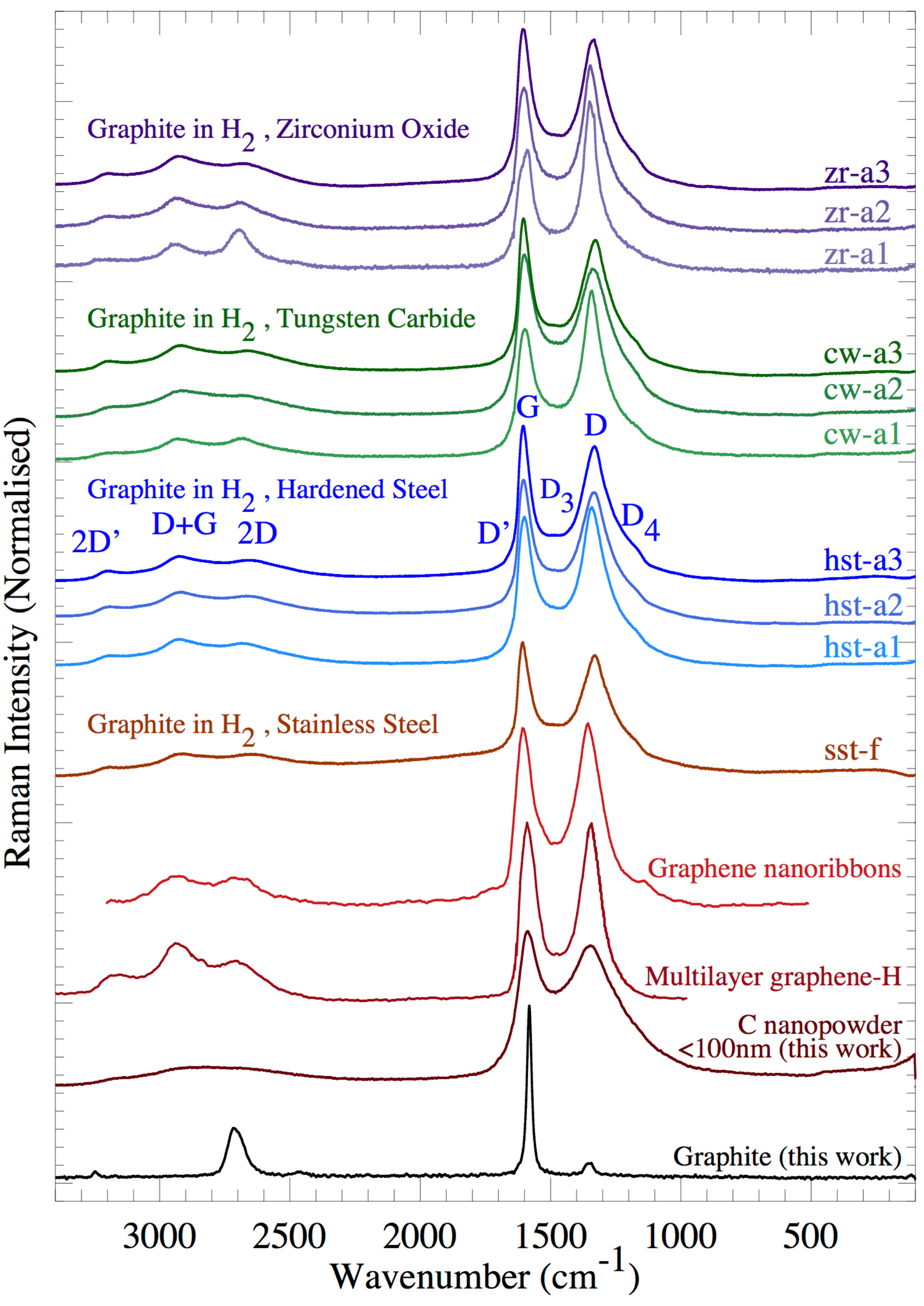}
\caption{Raman spectra of a sample set of planetary ball-milled graphite under a pressurised hydrogen atmosphere in function of milling time. The millings were performed in bowls made of stainless steel, hardened steel, tungsten carbide, and zirconium oxide. The Raman intensities were normalised. A graphite chunk and carbon nanopowder were measured for comparison. The spectra recorded for hydrogenated multi-layer graphene \citep{Ilyin2011} and graphene nanoribbons \citep{Jovanovic2014} are displayed.}
\label{fig:Raman}
\end{figure}
%
%
%~~~~~~~~~~~~~~~~~~~~~~~~~~~~~~~~~~~~~~~~~~~~~~~~~~~~~~~~~~~~~~~~~~~~~~~~~~~~~~~~~~~~~~~~~~~~~~~~~~~~~~~~~~~~     
\subsection{Infrared \& Raman measurements}
%infrared
Infrared measurements were performed by flattening the samples between two diamond windows.
Some of the samples were recorded with an In10 Bruker infrared microscope located on the SMIS synchrotron SOLEIL beam line. The infrared spot size was optimised to cover homogeneous flat parts of the samples (typically 10 to 200 $\mu$m projected aperture). The spectra were recorded with a sensitive cooled (MCT) detector covering the 4000-650 cm$^{-1}$ range, and, when possible, a room temperature detector (DTGS) to extend the (less sensitive) measurements down to 400 cm$^{-1}$. Other samples were measured using a Nicolet Continuum XL IR microscope located at the Laboratoire de Physique des 2 Infinis Ir\`ene Joliot-Curie and recorded with a sensitive cooled (MCT)
%\LEt{Reminder: please ensure & check throughout the paper that all acronyms and abbreviations are written out at first mention, followed by the abbreviation or acronym in parentheses (even if you have already introduced them in the Abstract). \textbf{\uline{After that please use only the abbreviation}}. Instruments, surveys, or facilities do not need an introduction in the Abstract. Please introduce these in the body of the paper unless they are known only by their acronym. See Sect. 5.2.4 of the Language Guide (https://www.aanda.org/for-authors/language-editing/1-introduction)  } 
detector covering the 4000-700 cm$^{-1}$ range and when possible another less sensitive extended (MCT) detector to cover down to about 500 cm$^{-1}$.\\
A Thermo Fisher DXR spectrometer located on the SMIS synchrotron SOLEIL beam line was used to record a Raman spectrum for the samples measured in the infrared, using a laser source at 532 nm at minimum power (100 $\mu$W on sample) to prevent any sample alteration. About ten scans with integration times of 20s each were co-added for each spectrum. The spot size of the Raman spectra is of the order of a micron.
%________________________________________________________
%~~~~~~~~~~~~~~~~~~~~~~~~~~~~~~~~~~~~~~~~~~~~~~~~~~~~~~~~~~~~~~~~~~~~~~~~~~~~~~~~~~~~~~~~~~~~~~~~~~~~~~~~~~~~     
\subsection{High-resolution transmission electron microscopy}

Milled sample extracts were diluted in pure ethanol solvent, and a drop was transferred to lacey carbon films supported by 3mm TEM grids. Once the solvent evaporated, the grids were mounted in a JEOL JEM-2011 HRTEM microscope operating at an acceleration voltage of 200 kV, from the technical platform managed by the Institut des Mat\'eriaux de Paris Centre (IMPC) from the CNRS and Sorbonne Universit\'e.
The microscope is equipped with 
a  GATAN system ORIUS SC100 (4008 $\times$ 2672 pixels) CCD camera.
Low magnifications (10,000 - 50,000 $\times$) were used for imaging the grain morphologies, and the inclusion of contamination by milling ball material chips that were incorporated in the samples during the milling process.
Higher magnification (100,000 - 800,000 $\times$) were used to visualise the profiles of the atomic planes parallel to the incidental
beam, and thus the evolution of the nanostructure. 
To minimise any overlap of the fringes, we selected the thinnest edges in the images at high magnification.
%
%~~~~~~~~~~~~~~~~~~~~~~~~~~~~~~~~~~~~~~~~~~~~~~~~~~~~~~~~~~~~~~~~~~~~~~~~~~~~~~~~~~~~~~~~~~~~~~~~~~~~~~~~~~~~     
\subsection{Elemental analysis}

The hydrogen and carbon contents of a subset of the produced samples were determined by the SGS company and are reported in Table~\ref{Table_summary}. They were measured after complete combustion in oxygen at 1150$^{o}$C in a helium flux of several mg of material, by sequentially analysing the resulting combustion gases by thermal conductivity.
        
%
%       FIGURE 5
%
\begin{figure*}%[tbhp]
\centering
\includegraphics[width=1.0\linewidth]{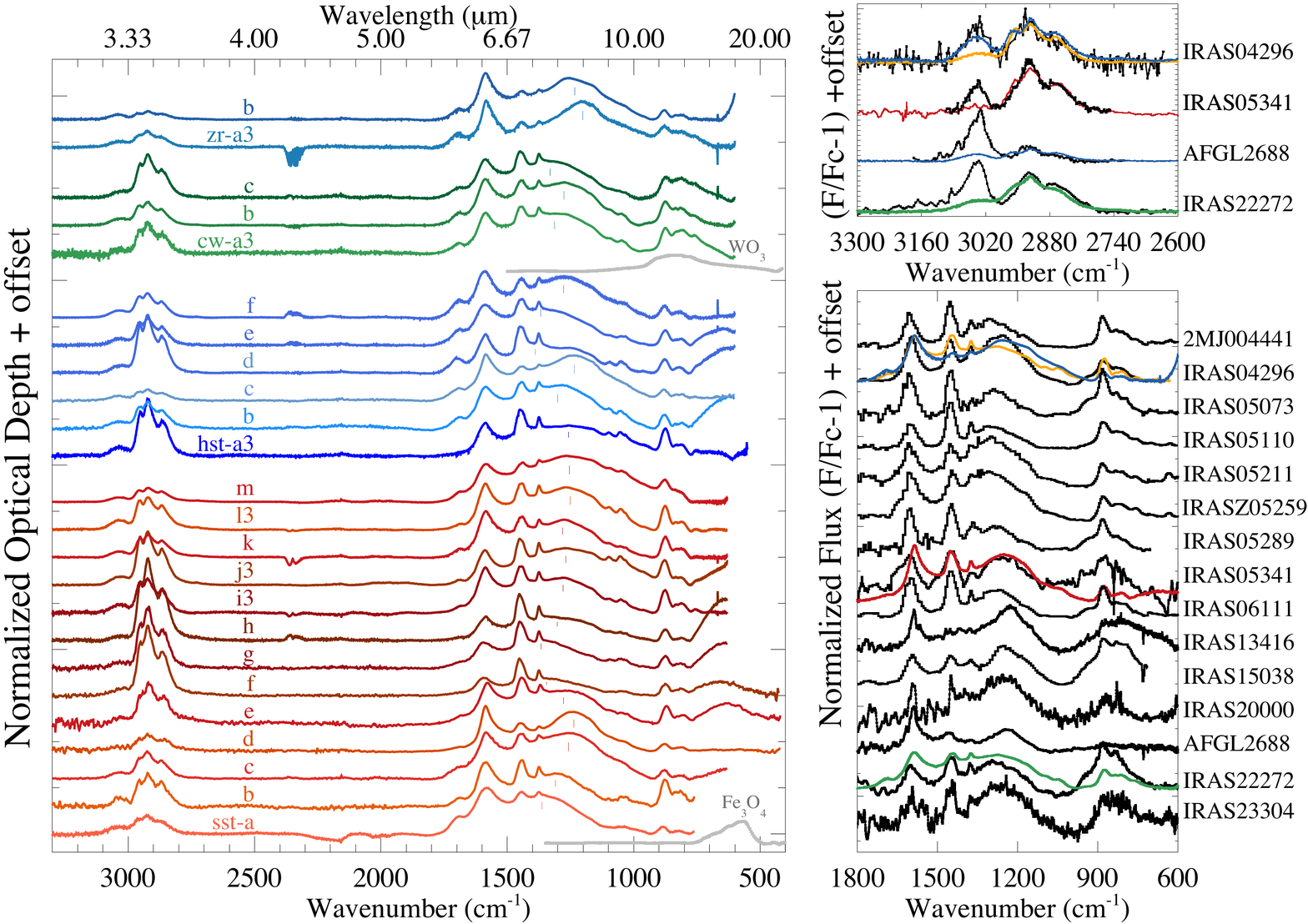}
\caption{Left panel: Infrared spectra of the mechanochemically produced interstellar analogues, baseline subtracted. The colours correspond to different milling materials (stainless steel: red; hardened steel: blue; tungsten carbide: green;  zirconium oxide: cyan). Infrared spectra of potential main contaminant for long-duration milling times are shown for clarity: iron oxide nanoparticles \citep[Fe$_3$O$_4$ spectrum from][]{Can2010} for stainless steel and hardened steel and tungsten oxide \citep[WO$_3$ spectrum from][]{Dawson2011}, expected to be marginal due to its high shear resistance, for tungsten carbide. The thin lines indicate the estimated position of the peak maximum for the broad massif. The nomenclature refers to the description in Table~\ref{Table_summary}.
Right panels: Astronomical observations for comparison to the analogue spectra. Top: Near-infrared literature spectra available for four out of all the analysed sources. Bottom: Mid-infrared Spitzer and ISO spectra of class $\mathcal{C}$ and $\mathcal{D}$ type sources analysed.
For the spectra including both near and mir spectra, a combination of analogue spectra are overlaid for comparison.}
\label{fig:IR_broyages}
\end{figure*}
%
%~~~~~~~~~~~~~~~~~~~~~~~~~~~~~~~~~~~~~~~~~~~~~~~~~~~~~~~~~~~~~~~~~~~~~~~~~~~~~~~~~~~~~~~~~~~~~~~~~~~~~~~~~~~~     

%
%       FIGURE 4
%
\begin{figure*}%[tbhp]
\centering
\includegraphics[width=1.0\linewidth]{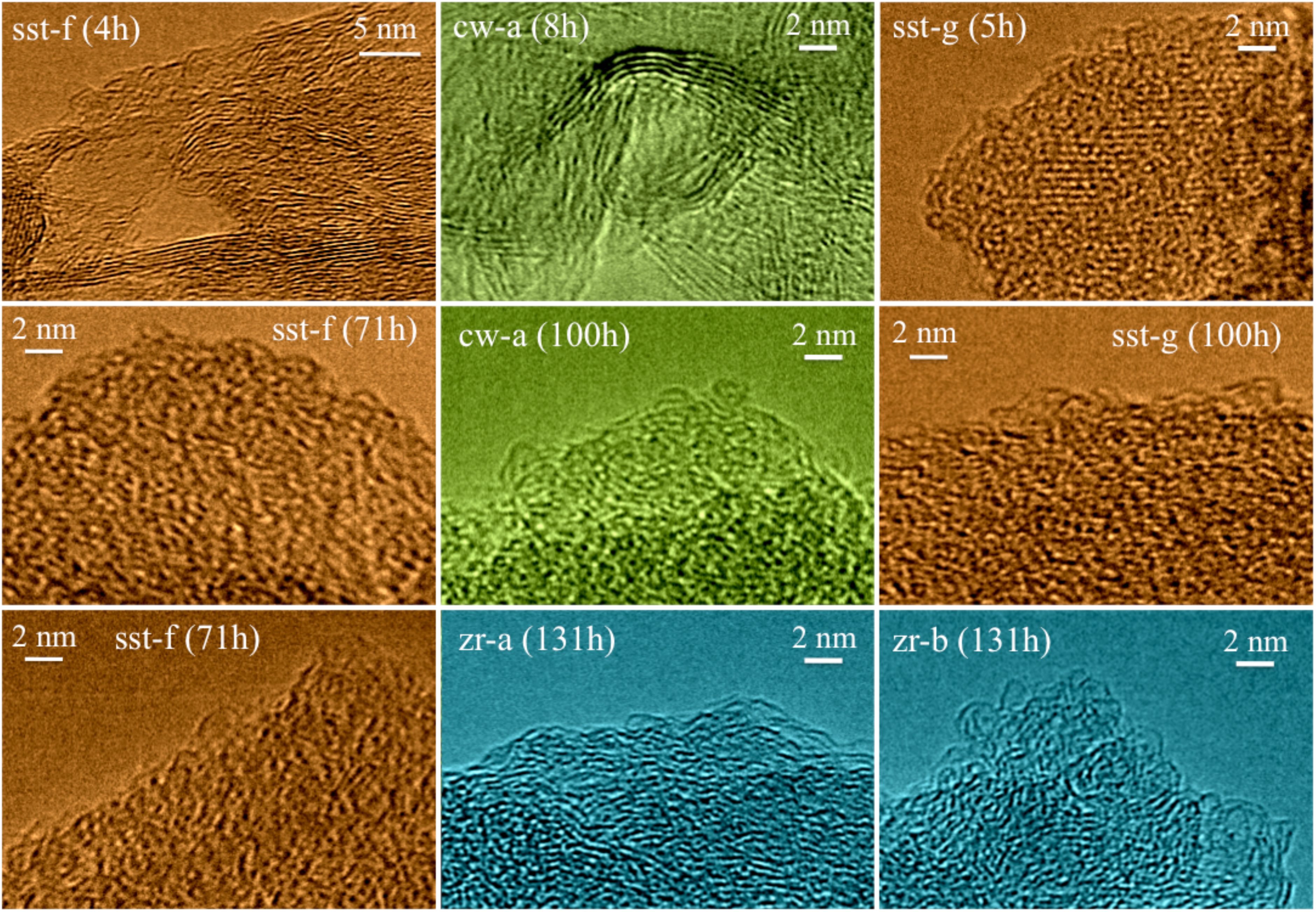}
\caption{High-resolution transmission electron microscope images for different ball-milled carbon precursors of this study and different mill materials. They are shown at large mechanochemical milling times performed in an hydrogen atmosphere. A few images are shown after a few hours of ball milling for comparison. The colour coding for the images is brown for stainless steel, green for tungsten carbide and light cyan for zirconium oxide experiments, respectively.}
\label{fig:HRTEM_general}
\end{figure*}
%
%
%
%       FIGURE 
%
\begin{figure}%[tbhp]
\centering
\includegraphics[width=1.0\linewidth]{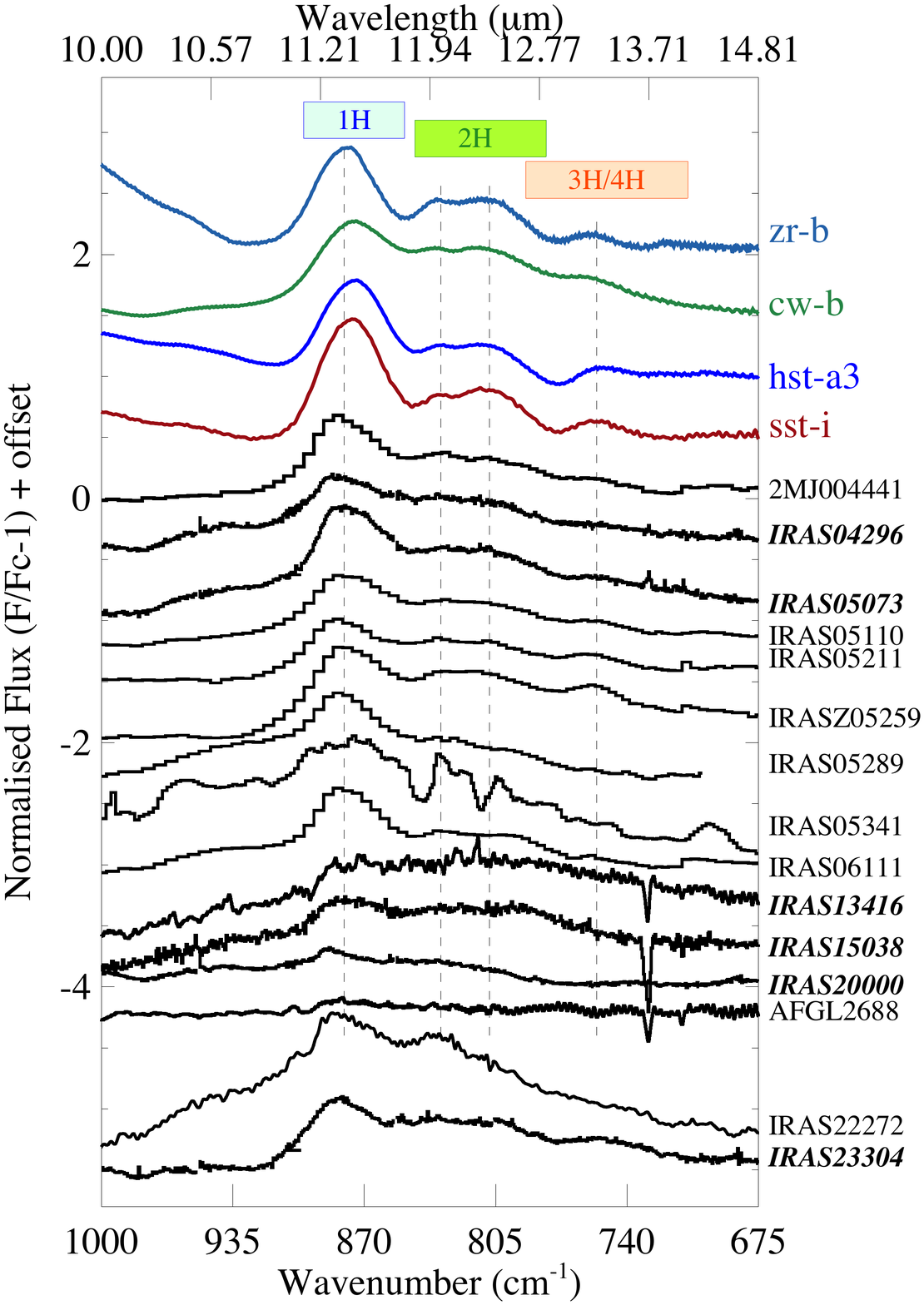}
\caption{Zoom on Spitzer and ISO spectra of class $\mathcal{C}$ and $\mathcal{D}$ type sources analysed in the out-of-plane aromatic CH mode's spectral region. The sources labelled in italics correspond to available Spitzer high-resolution spectra (R$\sim$600). Several analogues produced in different ball-milling environments are displayed above, for a direct comparison of the position of the observed bands and a range of measured relative intensities. The typical span of expected positions for the CH out of plane bands according to the underlying aromatic structure to which they are attached are given above (1H: solo; 2H: duo - bay, armchair; 3H: trio; 4H: quartet).}
\label{fig:oop}
\end{figure}
%
%~~~~~~~~~~~~~~~~~~~~~~~~~~~~~~~~~~~~~~~~~~~~~~~~~~~~~~~~~~~~~~~~~~~~~~~~~~~~~~~~~~~~~~~~~~~~~~~~~~~~~~~~~~~~     
\section{Results}

The infrared spectra of a sample set of the produced interstellar analogues by planetary ball milling graphite under a pressurised hydrogen atmosphere with various milling media (stainless steel, hardened steel, tungsten carbide, zirconium oxide) are shown in Fig.\ref{fig:IR}. Samples were extracted at different milling times to follow the evolution of the transformation. The spectra were continuum subtracted. Electron microscope images of minute amounts of the extracted samples at different milling times, for the hardened steel experiment, are shown in Fig.\ref{fig:HRTEM} and were recorded at various spatial scales down to the nanometre range. The images show large agglomerates of particles modified at the nanometre scale. Some coated steel nanoparticles are observed, embedded within the dominating carbon phase, ripped from the bowl and/or steel balls during the violent milling. The images show that the ordered graphite structure is affected even after the first milling phases, and any turbostratic structure totally disappeared at long milling times. Even in the absence of a hydrogen atmosphere, the mechanical milling is known to generate disordered carbon \citep[e.g.][]{Salver-Disma1999}. The addition of hydrogen to the mechanically produced carbon defects induces the saturation of reactive sites \citep[e.g.][]{Francke2005}.

Raman spectra of a sample set of the produced interstellar analogues were recorded and are shown in Fig.\ref{fig:Raman}. The pristine graphite used for the controlled atmosphere milling is also shown, as well as a carbon nanopowder spectrum, literature spectra of multi-layer graphene \citep{Ilyin2011} and graphene nanoribbons \citep{Jovanovic2014} for comparison. 
Different bands are observed. The G band ($\sim1580-1600$cm$^{-1}$), which is the so-called 'graphite' band, is in fact present in all sp$^2$ carbons. The D band ($\sim1350$cm$^{-1}$) is the Raman signature of disorder in sp$^2$ systems. The D' band ($\sim1620$cm$^{-1}$) is best observed in the zr-a1 spectrum where the G band displays an asymmetry that is clearly the sum of two bands.
A D$_3$ band ($\sim1500$cm$^{-1}$) fills the region between the D and G bands, attributed to the development of an amorphous carbon phase \cite[e.g.][]{Tallant1998, Sadezky2005}.
The D$_4$ band, which appears as a shoulder on the low wavenumber side of the D band, and which seems to grow with the milling time, can eventually be attributed to hydrocarbon or aliphatic moieties connected on graphitic basic structural units \citep[e.g.][]{Bokobza2015, Sadezky2005}.
The relatively intense and broad bumpy second order bands (2D', D+G, 2D) are indicative of the presence of numerous structural defects in the milled analogues, with the possible contribution at 2920 cm$^{-1}$ of sp3 CH$_x$ to the D+G band \citep{Ferrari2001}.

The infrared spectra of the many ball-milled precursor experiments, at long milling times, using different milling media and conditions are shown in the left panel of Fig.\ref{fig:IR_broyages}. 
High-resolution transmission electron microscope images for some of the samples are given in Fig.~\ref{fig:HRTEM_general} to follow the structural evolution of the samples under milling.
A comparison of absorption spectra with the infrared emission spectra of astronomical sources is shown in the right panels of Fig.\ref{fig:IR_broyages}. To perform such a comparison, the emission spectra of the astronomical sources already presented in Fig.\ref{fig:obs} were divided by an emission grey-body spectrum, which was estimated by fitting a spline on the continuum outside the bands, at about 3300, 2600, 1800, 1000, and  550 cm$^{-1}$. This is equivalent to assuming that the aromatic infrared emission band carriers emit at the same temperature as the underlying bands'
%\LEt{or bands' depending on if it's one or more bands... please do check this point throughout as I can't be sure which you mean, so I risk altering your meaning} 
continuum in this wavelength range. This step is necessary for a first-order 
comparison of the relative intensities to the one measured in absorption for the analogues in the laboratory. In particular, we note that when this emission correction is applied, the long wavelength polyaromatic CH out of plane modes (in the 700-900 cm$^{-1}$ range) appears comparatively weaker than the other vibrational modes and than what is observed in the raw astronomical emission spectra. The relative intensities are then closer to the one measured in the laboratory for the analogues.
A zoom in on the out-of-plane aromatic CH modes is presented in Fig.\ref{fig:oop}, together with representative spectra of the analogues in different bowl-milling media.
Several diagnostics are placed on these data. The integrated optical depth ratio of the CH$_{x={2,3}}$ modes in the stretching region from about 2800 to 2980~cm$^{-1}$, hereafter called 2920~cm$^{-1}$ to the corresponding deformation modes, $\mathcal{R}=$2920~cm$^{-1}$/(1460~cm$^{-1}$+1375~cm$^{-1}$), varies only slightly in the analogues produced with $\mathcal{R}=5.9\pm1.4$. 
The integrated absorption of the aromatic C=C band at 1600 cm$^{-1}$, composite aliphatic CH$_{x={2,3}}$ band at about 1460cm$^{-1}$,  and CH$_{3}$ band at about 1375cm$^{-1}$ are evaluated, both in the analogues and the astronomical normalised spectra. These bands give access to the aliphatic C-H to aromatic C=C content of the samples and sources. The analysis focuses on the 1600 to 1000 cm$^{-1}$ range. This range is chosen intentionally to build analytical ratio for several reasons. Firstly, as they are close in energy, their relative value in emission is expected to be least affected by temperature in the case of comparisons to astronomical spectra. Indeed, as can be seen from Fig~\ref{fig:obs}, the three-micron continuum slope can be opposite (e.g. IRAS04296) to that of the continuum emission at a longer wavelength. It suggests that the emission arising at such wavelengths may come from an inner part, from a smaller distance to the astronomical source, and this spectral range cannot easily be compared to the mid infrared emitted from another region without precise knowledge of the source environment's geometry.
The large band evolving between 1170 and 1350 cm$^{-1}$ is also interesting, as it is an important diagnostic of the spectral evolution of the AIB's astronomical classes. Its potential assignments are discussed in the next section. The position of the maximum of this band in both observations and laboratory analogues are measured.

\begin{figure}%[tbhp]
\centering
\includegraphics[width=1.0\linewidth]{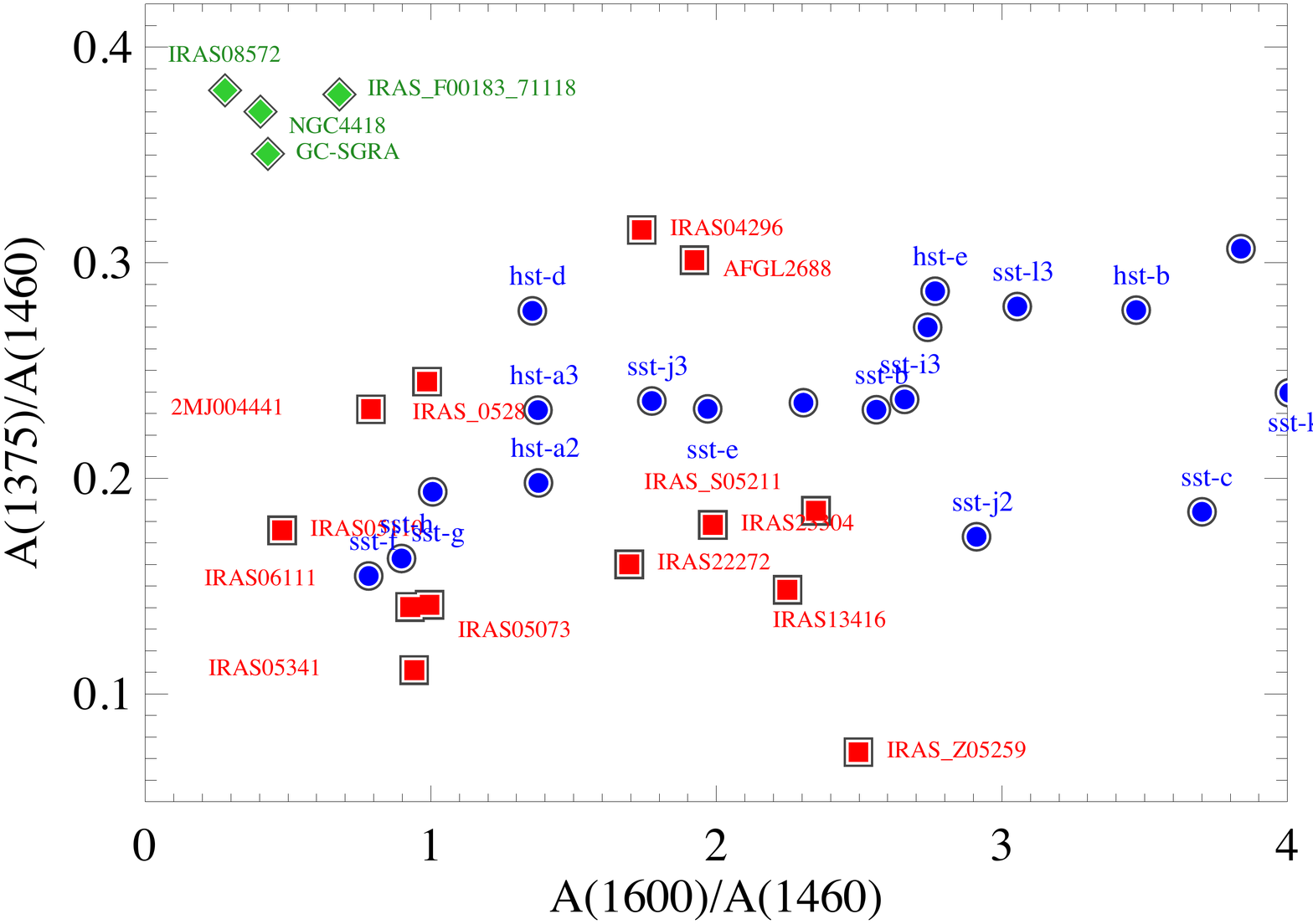}
\caption{Comparison between astronomical spectra integrated-intensity ratios of aromatic C=C band at 1600 cm$^{-1}$ to aliphatic CH$_x$ band at 1460cm$^{-1}$ according to the aliphatic CH$_x$ 1375cm$^{-1}$/1460cm$^{-1}$ ratio. 
The green diamonds correspond to the diffuse interstellar medium a-C:H absorptions. 
The red squares correspond to the class $\mathcal{C}$ and $\mathcal{D}$ sources in this study. 
The blue dots are the same ratios measured on the infrared spectra of the produced analogues.}
\label{fig:rapports}
\end{figure}

\begin{figure}%[tbhp]
\centering
\includegraphics[width=1.0\linewidth]{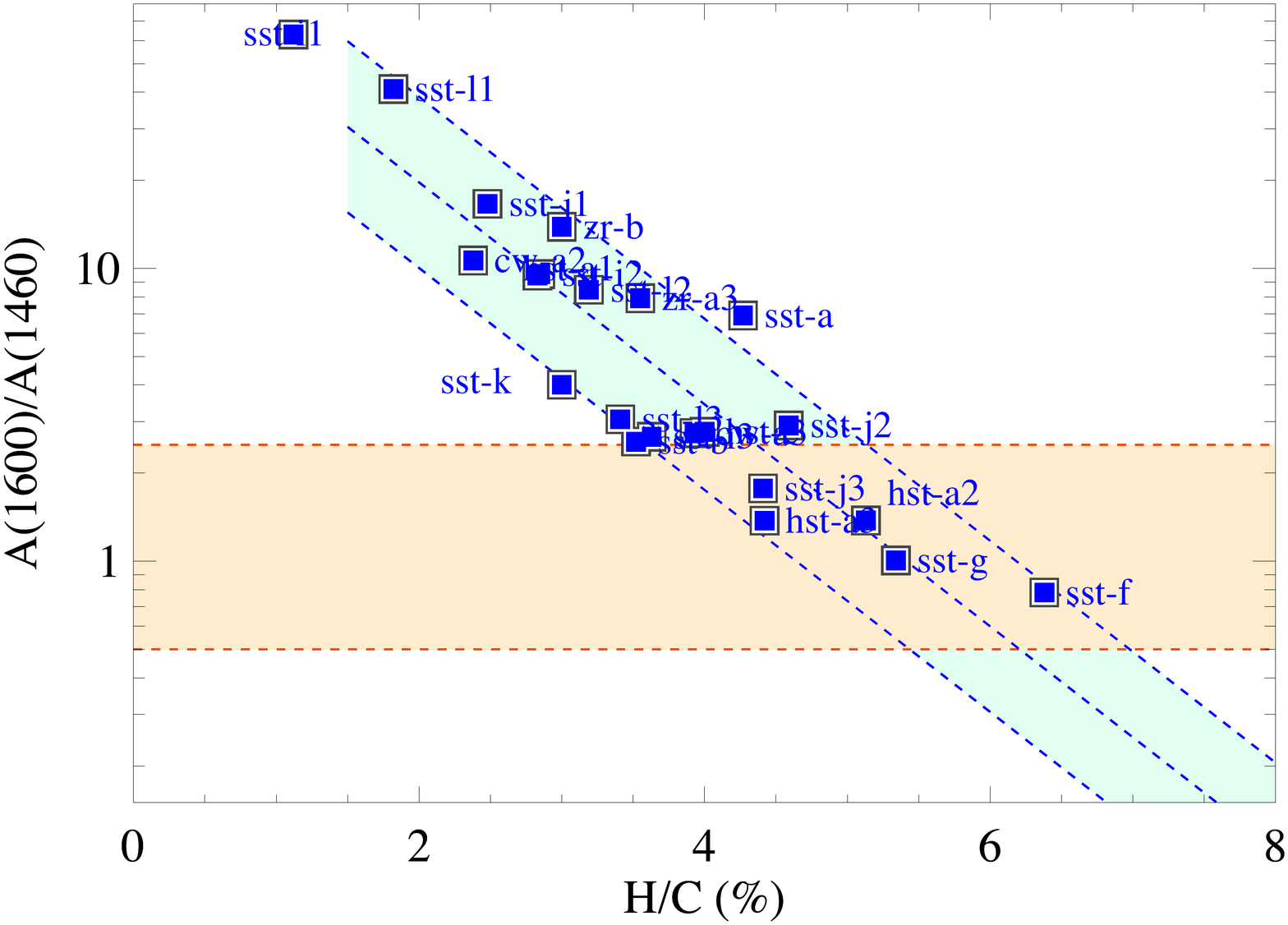}
\caption{Comparison between analogue spectra integrated-intensity ratios of aromatic C=C band at 1600 cm$^{-1}$, and aliphatic CH$_x$ band at 1460 cm$^{-1}$ as a function of the elemental hydrogen to carbon ratio measured independently. The blue coloured region encloses the confidence interval of the correlation. The orange coloured region corresponds to the observed range for astronomical observations.}
\label{fig:1600_1460_versus_h_sur_c}
\end{figure}

\begin{figure}%[tbhp]
\centering
\includegraphics[width=1.0\linewidth]{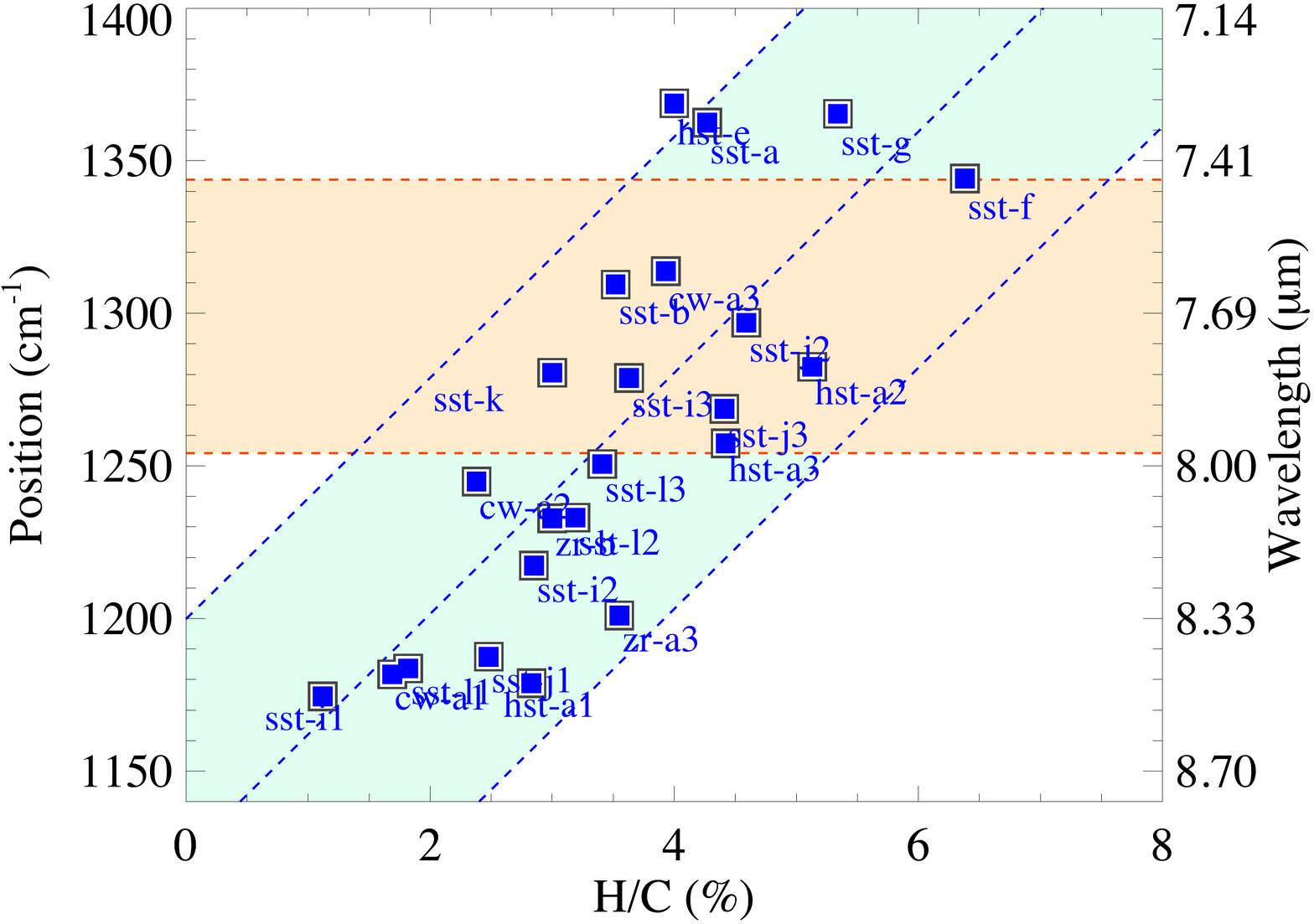}
\caption{Evolution of the '1300cm$^{-1'}$ peak maximum position in analogue spectra displayed according to  the elemental hydrogen to carbon ratio measured independently. The blue coloured region encloses the confidence interval of the correlation. The orange coloured region corresponds to the observed range for the astronomical observations analysed in this article.}
\label{fig:1300_versus_h_sur_c}
\end{figure}

From these evaluations, the 1375cm$^{-1}$ CH$_{x={2,3}}$ to 1460cm$^{-1}$ CH$_{x={2,3}}$ ratio is plotted in function of the 1600 cm$^{-1}$ C=C to 1460cm$^{-1}$ CH$_{x={2,3}}$ ratio in Fig.~\ref{fig:rapports}, for the class $\mathcal{C}$ and $\mathcal{D}$ sources, a-C:H diffuse ISM sources for comparison, and the mechanochemically produced analogues. In the dust grain analogues, the 1600 cm$^{-1}$ C=C to 1460cm$^{-1}$ CH$_{x={2,3}}$ ratio is plotted in function of the measured elemental H/C ratio in Fig.~\ref{fig:1600_1460_versus_h_sur_c}. The range for the considered observations is over-plotted for comparison. In addition, the '1300' cm$^{-1}$ ("7.7$\mu$m") massif peak maximum position is measured in the analogues and observations. Its position with regard to the elemental H/C ratio is displayed in Fig.~\ref{fig:1300_versus_h_sur_c}.


%~~~~~~~~~~~~~~~~~~~~~~~~~~~~~~~~~~~~~~~~~~~~~~~~~~~~~~~~~~~~~~~~~~~~~~~~~~~~~~~~~~~~~~~~~~~~~~~~~~~~~~~~~~~~     
\section{Discussion}

During the milling process of carbonaceous precursors, the energies are high enough to shatter and break particles by high-frequency impact and intensive frictions, reducing their sizes down to hundreds of nanometres, and also profoundly affecting the samples at the atomic bonds level. 
In the presence of a hydrogen atmosphere pressurisation of the bowls, in addition to structural modifications, chemical reactions proceed, and hydrogenation of the produced particles' reactive sites further modifies the dust particles.
This process is particularly attractive for forming astrophysical analogues, as it might circumvent one of the issues when producing laboratory analogues, namely the fact that in many occurrences, the analogue modifications are produced at a macroscopic scale. This high-impact/shattering milling process produces and hydrogenates defects very locally in the ball-milled carbonaceous dust particles, inducing multi-scale defects.

The spectra of the produced analogues show marked similarities with astronomical observations. 
In particular, the analogues are able to provide a satisfactory spectral match to the 1300~cm$^{-1}$ massif, together with large aliphatic deformation mode ($\sim$1460-1375~cm$^{-1}$) to aromatic C=C stretching mode ($\sim$1600~cm$^{-1}$) intensity ratios. Some of the bands are, however, still slightly broader than the astronomically observed ones. The C=C mode shows an asymmetric profile warped toward longer wavelengths as observed in astronomical spectra.

The out-of-plane CH modes in the produced analogues, peaking at about 880~cm$^{-1}$ and attributed to isolated solo H (1H), at 832 and 808~cm$^{-1}$ and attributed to duos H (2H), at 755 cm$^{-1}$ and attributed to trios or quartets H (3H,4H), are close to the astronomically observed peak positions and in agreement with their relative intensities.
%\LEt{I'm not sure to understand this first sentence. Up to 880 cm, is one of the words a verb? Please review.  }
The prevalence of the aromatic solo H out of plane mode at about 880 cm$^{-1}$ in class $\mathcal{C}$ to $\mathcal{D}$ astronomical spectra is  reproduced. This band possesses an intrinsically larger band strength than the other out-of-plane modes \citep[e.g.][]{Bauschlicher2009}, by a factor up to about 4. If the integrated intensities in the duo modes (832, 808~cm$^{-1}$) are lower, the duos may contribute equally to the underlying structure. By contrast, the low intensities in the trio/quartet bands region of the spectrum supports the idea of hydrogen bonded to isolated or peripheral hydrogen (1H), or (2H) structures, as well as a rather low hydrogen coverage in the produced analogues. 

There are ample discussions and research regarding the relation between the polyaromatic structure, intensity, and position of the CH the out-of-plane bands in the literature, \citep[e.g.][]{Bauschlicher2009, Candian2014, Russo2014, Tommasini2016, Sasaki2018, Ricca2019}. For the 2H structure, the position can be typically associated with zigzag, armchair, or bay configurations.

The analogues' relative intensities fit the lower energy region of the spectra well. The $\rm6.2\,\mu m/6.85\,\mu m$ ($\rm1600\,cm^{-1}/1460\,cm^{-1}$) ratio that fit the observations is lower than what is expected when compared to the CH stretching-mode-region observations. In the CH stretching mode region, the aromatic ($\rm\sim3040\,cm^{-1}$) to aliphatic ($\rm\sim2760-2980\,cm^{-1}$) CH bands intensities are higher, by a factor of a few units, in the sources observed (IRAS 04296-3429, IRAS 05341+0852, AFGL2688, IRAS 22272+5435, see upper-right panel of Fig.~\ref{fig:IR_broyages}) than in the analogues. It suggests that the emission in this spectral range may, as mentioned above, arise in regions closer to the central source, with species possessing a slightly higher aromatic CH content. %The 

The comparison between the observations' intensity ratios of the aromatic C=C band at 1600 cm$^{-1}$ to aliphatic bands at 1460 cm$^{-1}$, versus the 1375-to-1460 cm$^{-1}$ ratio immediately shows that these ratios are definitely different from the Galactic hydrogenated amorphous component observed in absorption in the diffuse interstellar medium (Fig.\ref{fig:rapports}). On the contrary, these ratios agree fairly well with the analogues produced in this study. The lower 1375-to-1460 cm$^{-1}$ ratio is an interesting criterion characterising a low(er) methyl-to-methylene ratio in the aliphatic CH inserted in the structure (although the exact relative intensities could be affected by a high stress in the material). CH$_2$ may  bridge faults in the aromatic structure. 

The ratio of the aromatic C=C band at 1600 cm$^{-1}$ to the aliphatic CH$_{x={2,3}}$ band at 1460 cm$^{-1}$ (Fig.\ref{fig:1600_1460_versus_h_sur_c}), does not evolve linearly with the elemental hydrogen-to-carbon ratio (measured independently). In the milling process it implies that the first hydrogen atoms chemically attached form mainly aromatic C-H bonds at early times. The aliphatic C-Hs are formed in the next chemical attack steps, probably at the carbon sites destabilised by the first inserted hydrogens. 

The amount of material produced by this method made it possible to perform an elemental analysis to determine the H/C ratio for a subset of the analogues. A comparison between the 1600 /1460 cm$^{-1}$ ratio in the analogues versus their H/C content is presented in Fig. \ref{fig:1600_1460_versus_h_sur_c}. The corresponding range measured on the astronomical spectra is shown in light blue. The agreement between the spectra and the observations is obtained for an hydrogen to carbon ratio of 5$\pm$2\%, setting a strong constraint.

The emission massif in the 1170 to 1300 cm$^{-1}$ range (from about 8.6 up to 7.6~$\mu$m) is one of the spectral structures that is among the most specific for the AIB carriers, with a marked evolution from the class $\mathcal{A}$ to $\mathcal{D}$ \citep[e.g.][]{vanDiedenhoven2004, Matsuura2014}. In the produced analogues, the vibrational modes absorbing in this range can be associated with geometrical distortions of the C-C skeleton \citep[e.g.][and references therein]{Centrone2005, Lazzarini2016}, in addition to potential moderate C-O-C oxygen contamination mode contributions in the wing of the profile, as they peak mainly in the lower part of this wavenumber range. The position of the 1300 cm$^{-1}$ massif thus puts a strong constraint on the internal structure or carbon based skeleton of the AIBs carriers. In the spectral evolution shown in Fig.\ref{fig:IR} this massif evolves gradually from about 1200 cm$^{-1}$ after a few hours to 1270-1350 cm$^{-1}$ at large milling times. Its position is related to the onset of important curvature imposed by the production of nanoscale structures as sketched in Fig~\ref{fig:experience}. The HRTEM images of the ball-milled precursor at long milling time displayed in Fig.\ref{fig:HRTEM_general} show the ubiquitous nature of such warped and round curvatures with radii as small as 1 to 2 nm. The position of the massif is reported in function of the H/C ratio in Fig.~\ref{fig:1300_versus_h_sur_c}. If a rather large dispersion still exists, this figure shows a positive correlation trend of the position with the H/C content in the analogues. Thus, the chemical insertion of hydrogen influences (or betrays) the underlying carbon network structure.

\section{Conclusions}
Mechanochemistry represents a pertinent methodology to produce carbonaceous interstellar analogues, starting from pure carbon-based precursors immersed in a reactive hydrogen atmosphere. The impact and shear milling at high energy produces local defects down to nanometre level and leads to a partial hydrogenation of resulting polyaromatic interlinked carbon structures.
The main outcomes of this work are:
\begin{itemize}
\item These mechanochemical analogues successfully reproduce the overall spectra of  polyaromatic infrared  emission bands observed towards some astrophysical class $\mathcal{C}$ and $\mathcal{D}$ type sources, including the poorly assessed and elusive broad emission band peaking in the 7.6-8.6~\si{\mu\meter} ($\rm\sim1170-1300~cm^{-1}$) range, involving CC stretching and deformation modes.
\item The position of the 7.6-8.6~\si{\mu\meter} band shows a trend related to the hydrogen to carbon content of the analogues produced, measured independently. 
\item A correlation is observed between the intensity ratio of the C=C stretching at about 6.2-6.3~\si{\mu\meter} ($\rm\sim1600~cm^{-1}$)  and aliphatic C-H deformation at about 6.85~\si{\mu\meter}, ($\rm\sim1460~cm^{-1}$) and the hydrogen to carbon content. 
\item Transposed to astrophysical observations, the agreement between this ratio measured in analogues and observations in the mid-infrared range would imply the need for only about 3\%$\leq$H/C$\leq$7\% in the polyaromatic-emitting structure; this represents a relatively low hydrogen content given the reducing H-rich astrophysical media, to explain the observed spectra.%
%\LEt{Could you please add a brief paragraph to sum up after your bullet points. Thank you.} %
%
\end{itemize}
The mechanochemical approach to synthesise astrophysical carbonaceous dust particles, under an hydrogen atmosphere, is promising. It provides an adequate molecular structure for the reproduction of astronomical spectra, including the mid infrared bands. It enlightens us about the several percents hydrogen-to-carbon ratio in mixed aliphatic and aromatic carbonaceous dust required to explain the AIB observed emissions from
class $\mathcal{C}$ and $\mathcal{D}$ sources.

%-------------------------------------------------------------------

\begin{acknowledgements}
      This work was supported by the INTEGRITY Project, funded by the Domaine d'Int\'er\^et Majeur ACAV, labelled by the \^Ile-de-France region, the P2IO LabEx program: "Evolution de la mati\`ere du milieu interstellaire aux exoplan\`etes avec le JWST", the ANR COMETOR project, grant ANR-18-CE31-0011 of the French Agence Nationale de la Recherche, and the Programme National "Physique et Chimie du Milieu Interstellaire" (PCMI) of CNRS/INSU with INC/INP co-funded by CEA and CNES. The Combined Atlas of Sources with Spitzer IRS Spectra (CASSIS) is a product of the IRS instrument team, supported by NASA and JPL. CASSIS is supported by the "Programme National de Physique Stellaire" (PNPS) of CNRS/INSU co-funded by CEA and CNES and through the "Programme National Physique et Chimie du Milieu Interstellaire" (PCMI) of CNRS/INSU with INC/INP co-funded by CEA and CNES. 
       Some scientific results reported in this article are based on observations made by the Infrared Space Observatory (ISO), an ESA science mission, and retrieved using the ISO database.
      We would like to thank J. Guigand, J.-P. Dugal and H. Bauduin for their support in the mechanical design modifications of the experiment used in this study. 
      We would like to acknowledge the anonymous referee for the positive feedback on the article.
\end{acknowledgements}

%-------------------------------------------------------------------

%
\end{document}